\documentclass[review]{elsarticle}

\usepackage{subfigure}

\journal{Journal of \LaTeX\ Templates}

\begin{document}

\begin{frontmatter}

\title{FPGA-based real-time autocorrelator and its application in dynamic light scattering}
\tnotetext[mytitlenote]{Corresponding author.
}

%% Group authors per affiliation:
\author{Akhmarzhan Islambek}
\author{Kecheng Yang\fnref{myfootnote}}
\author{Wei Li}
\author{Kai Li}
\address{\centering School of optical and electronic information, Huazhong University of Science and Technology, Luoyu Road 1037,Wuhan, China}
\fntext[myfootnote]{\textit{E-mail address} 
kcyang@hust.edu.cn (Kecheng Yang)}

\begin{abstract}
Digital correlators play a significant role in dynamic light scattering (DLS) technology, which characterizes particle size distribution. We present a field programable gate array (FPGA)-based digital correlator that can be applied to process DLS data. To satisfy the DLS requirements in the FPGA logic with limited resources, a multiple lag time period (multi-$\tau$) method is employed that does not require storing the full dataset in memory. Moreover, the device directly accepts the transistor-transistor logic (TTL) signal from the photon counting detector by measuring the time intervals between photon events and calculates the autocorrelation functions in real time. Furthermore, we derive estimates for the error arising from the use of the multi-τ correlator. We implement all the necessary operations in a single Xilinx FPGA chip with a lag time from 10 ns to 45 min, including a highly optimized photon counter.
\end{abstract}

\begin{keyword}
\texttt dynamic light scattering, optical signal processing, single photon counting, digital correlator, FPGA design, VHDL  
\MSC[2010] 00-01\sep  99-00
\end{keyword}

\end{frontmatter}

\section{Introduction}

Dynamic light scattering (DLS) is used to measure the size of nanoparticles. With this method, it is possible to determine the diffusion coefficient of dispersed particles in a liquid by measuring fluctuations in the intensity of scattered light using the correlation function. A digital correlator is a device capable of calculating correlation functions in a DLS experiment by digitally processing the signal from a photon detector in the form of a pulse stream. Commercial digital correlators provided by Laser Vertriebsgesellschaft mbH (ALV) (Langen, Germany), Brookhaven Instruments (Holtsville, New York), and Correlator.com (Bridge-water, New Jersey) have good performance with a high dynamic range. However, these digital signal processor (DSP) devices have the disadvantages of inflexibility, high cost, and complexity of implementation, and it is worth considering alternative complementary approaches for ideal measurements. Currently, there are various types of customized digital correlators known for their implementation of this technique \cite{Engels99, Magatti2001,Schaub2012, Buchholz2012}. 
In DLS experiments, the real-time computation of correlation functions has to be handled over a large lag time period range with nanosecond resolution, which is associated with the concept of dynamic range. At the same time, the sampling time interval frequency for counting photoelectrons should be no more than 10 ns. To meet the required sampling rate, we implement a high-speed counter that operates at a frequency of 800 MHz. With the help of the multi-$\tau$ ($\tau$ is lag time period) technique, it is possible to balance the trade-off of resolution and dynamic range. Thus, the correlator is very efficient in terms of performance. Data analysis is initiated only after the data collection is complete. Our aim in this paper is to perform measurements with such conflicting criteria on an effectively implemented hardware correlator system.

\section{ Correlation theory and correlator architecture}

With a DLS setup \cite{HUST}, the autocorrelation function of intensity scattered by micro-particles is used to retrieve their particle size distribution \cite{Berne} through some algorithm such as CONTIN \cite{Provencher, ProvencherCON}, Bayesian method\cite{Clementi}, regularization method \cite{Twomey,Buttgereit}, neural network method \cite{Gugliotta} and etc. The autocorrelation function \textit{g($\tau$)} of scattered light expresses the relationship between a function (signal) $\textit{I}_1$ and its shifted copy $\textit{I}_2$ with the time interval $\tau$.

\begin{equation} \label{eq1}
g(\tau)=\langle I_1(t)I_2(t)\rangle=\frac{1}{N}\int_{0}^{N}I_1(t)I_2(t)dt
\end{equation}
                          
where \textit{N} is the total realization time. 

The basic operation principle of a digital correlator is to calculate the autocorrelation function at certain lag times $\tau_j$. In a discrete system, the autocorrelation function is expressed as:

 \begin{equation} \label{eq2}
g(\tau_j)=\frac{1}{N-j}\sum_{i=1}^{N-j} t(i)t(i+j)
\end{equation}
                            
where \textit{j} is the index of the correlation channel (\textit{j} = 0,1,2,..., \textit{J} -1), the parameter \textit{t(i)} is the photon arrival time of the \textit{i}-th sample (detected event), and \textit{N} is the total number of samples measured during the experiment. For the entire correlation system, $\tau_j (\tau_j = j \ast \delta_t )$ has parallel but different values $\tau_j : \tau_{min} < j \ast \delta_t < \tau_{max}$ to obtain the discrete approximation in real time. Since $\delta_t$ is the sampling time produced by delay element $\delta$, its interval is derived from the clock frequency \textit{f} of the \textit{j}-th channel $\textit{f}_{ch(j)}$ .

Dynamic range is a key parameter of a correlator that mainly depends on the number of correlation channels \cite{Jakob2007}. A linear correlator is a system consisting of a chain of correlation channels. If $\textit{f}_{ch(0)} = f_{ch(J-1)}$ in a linear correlator, then to achieve the desired dynamic range, a large number of correlation channels must be used, which is not feasible for hardware with limited resources. A more appropriate algorithm for a digital correlator is the multi-$\tau$ correlation technique, as described previously \cite{Schatzel1985, Magatti2003}, which allows increasing the dynamic range without increasing the number of correlation channels. The disadvantage of this technique is that averaging introduces a systematic error. Therefore, the data must be normalized by a symmetric normalization procedure \cite{Schatzel1988}.

Eventually, our photon correlator, implemented with the multi-$\tau$ technique, has $\textit{S}$ = 35 correlator blocks, each consisting of $\textit{P}$ = 8 equally divided correlation channels, except that the first block has $\textit{P}$ = 16 correlation channels. These parameters are set in this order to achieve high accuracy \cite{Magatti2001,Schatzel1990}. The sampling time interval for the first block is $\delta_{t(0)}$=10 ns, whereas it is increases with the factoor of $\textit{n}$ = 2 for every next block and reaches $\delta_{t(S-1)} \approx≈$ 2.9 mins in the last block. The entire correlation function is calculated internally using 288 correlation channels.

\section{ System design and its implementation}

\subsection{ Correlator structure and process control}

The design of our hardware photon correlator implemented on a general-purpose Nexys Video evaluation board (Digilent, USA) equipped with Artix 7 FPGA chip (XC7A200T) as illustrated in Fig.1.

As shown in Fig.1, all the functionally interdependent design components (the counter, autocorrelation unit, random access memory (RAM), multiplexer (MUX), soft processing core, local memory, universal asynchronous receiver-transmitter (UART)) are supplied with reset and clock synchronization process control signals from the same sources (reset system and clock configurator) simultaneously, as indicated by the arrows (yellow). We set the system clock configurator to generate the system clock frequency $\textit{f}_{sys}$ with 100 MHz.

In the correlator unit (overall correlator blocks), in addition to reset and clock synchronization signals, there are four more process control signals: clock enable, clear, start and stop.

According to the structure of the multi-$\tau$ correlator, the first block clock frequency $\textit{f}_{blk(0)}$ is constant; thus, $\textit{f}_{blk(0)} = \textit{f}_{sys}$, and from $\textit{f}_{blk(1)}$ to $\textit{f}_{blk(S-1)}$, the clock must be generated with different frequencies, as demonstrated in Fig. 2. The clock enable signal $\textit{e}_s$ is connected to the clock cycle counter $\textit{c}_s$ , where $\textit{s}$ is the index of the correlator block ($\textit{s} = 0,1,2,3,..., S -1$). The clock cycle counter $\textit{c}_s$ range is from 1 to $\textit{e}_s$. Thus, whenever the clock cycle counter $\textit{c}_s$ output value is equal to $\textit{e}_s$, the clock enable signal is activated and remains inactive until looping returns this condition again. This method decreases the consumption of the flip-flops, thus increasing the performance and reducing the power supply in the design.

Another process control signal of the correlator unit is a clear signal used to reset the accumulator data in the correlation channels. The capacity of the accumulator is 64-bit, and all the accumulator’s data in the autocorrelator unit must be read out to the internal block RAM for data storage with a 20-s cycle to avoid overflow.

In our design, the correlator system is controlled by finite state machines (FSM). In turn, the FSM send the start and stop signals to the autocorrelator unit (see Fig. 2). In general, there are four states optimizing the correlator processing system: “1”, the initial idle state; “2”, the ready for data processing state; “3”, the data processing state; and “4”, the end operation.

In FPGAs, interconnect systems are used to tie processors to peripherals. The Advanced Extensible Interface (AXI) interconnect makes a connection between the processing system and programmable logic in the stream terminology by mapping to the processor memory, and it also provides a point-to-point unidirectional connection from the soft processing core to the UART peripheral interface for sending the correlator results.

To print the correlator results, there are two nested cycles. In the first cycle, the processing core selects a memory (RAM) row to print, and in the second cycle, the processor selects a memory column to print. The embedded software library of the processing core is located in local memory (see Fig.1.). The functionality of the soft processing core mainly depends on the machine states of the correlator and acts in accordance with its state changes, except for the final state. In the “4” state, the processor core starts to continuously send a data stream containing the autocorrelation results to a personal computer (PC) via a universal serial port (USB). Thus, the PC interface stores all the received data in a dedicated file.

 The data flow steps through the system represented by sequence number are shown in Fig.1.

\subsection{Photon detection/counting module}

This module is used to connect the correlator with the photon detector and guarantee adequate photon detection efficiency. In our case, the scattered light is measured by a photon detector (Hamamatsu H10682-210) at four different scattering angles ($\theta$ =$15^\circ$, $30^\circ$, $45^\circ$, and $60^\circ$), and a neutral density (ND) filter is placed before the incident light to avoid detector saturation. The incident light is filtered by a suitable ND filter to guarantee that the detector obtains a count per second (CPS) of approximately $5\times10^6$.

Another barrier is the power supply difference between the photon detector (5 V) and the FPGA board (3.3 V). A single buffer (NC7SZ125) with three state outputs on a footprint board (SOT-23) is mounted as a level converter. In this way, we deliver photon pulses to the FPGA peripheral module (PMOD) interface with the 10 ns output pulse width of the photon detector.

The time-correlated single photon counting (TCSPC) technique is effective due to its ability to count detector pulses within defined time intervals. Since our correlator is designed for data acquisition with a TCSPC technique, we use a fast gate in the front of the counter, which samples 1-bit data at 800 MHz. Then, with a deserializer the signal converts to 8-bit and sends to a counter which outputs the data with 100 MHz. The operation principle of counter based on counting the number of clock cycles (running with a frequency of 800 MHz) between two events in a data set. Fig. 3 presents an example simulation waveform of the counter module.

According to the principle of operation of the counter, it takes the time interval of the clock cycle of 1.25 ns as 1 ns, which leads to the fact that the output value of the counter differs from its initial value. Then, to restore the actual data, the counter output data must be multiplied by 1.25 (10 ns/8). Fractional number multiplication in hardware multipliers has the problem of losing data accuracy. To avoid this problem, accumulated correlation functions are multiplied by $1.25^2$ when they are collected by the PC.

\section{Experimental results}

The correlator designed in section 3 was mounted in a DLS setup designed by our research group \cite{HUST} and was tested with a couple of polystyrene spheres produced by Suzhou Nanomicro Technology Company. The theoretical diameters of the particle samples were 240, 360, 530 and 805 nm with an accuracy on the order of 1 \% or less. In the preparation of the solution, we added one drop (concentration of 1\% in 10 ml) of latex to dust-free water. The prepared samples were placed in a 9-mm diameter cylindrical quartz cell.
The parameters we used for the DLS data simulation and experiment are as follows:  temperature \textit{T} = 298.15 K, viscosity $\eta$ = viscosity 0.89 mPas, and a linearly polarized diode pumped laser (CrystaLaser) with wavelength $\lambda$ = 532 nm and a corresponding particle refractive index is \textit{n} = 1.59. The laser beam illuminated the beads, and the scattered light was collected by detectors placed at scattering angles of ($\theta$ =$15^\circ$, $30^\circ$, $45^\circ$, and $60^\circ$).

In Fig. 4 we present analyses of the experimental samples obtained with our correlator, which demonstrate the movement of particles at the four different angles. The correlation function is normalized by a symmetric normalization procedure [21].

 Fig. 5 presents the behaviour of particles with different sizes at the same angle ($30^\circ$). To predict the deviations of the actual values from the current values we fit the autocorrelation curve according to equation:

\begin{equation} \label{eq4}
g(\tau)=B+\beta\exp(-\Gamma\tau)
\end{equation}

where the the baseline \textit{B}, the amplitude $\beta$, and decay rate \textit{$\Gamma$} are the fitting parameters.

According to the principles of DLS, the decay rate of the generated autocorrelation curve is in the range of ~100 ns - 45 min (10 to 100 ns is the time required for the device to warm up), as shown in Fig. 5.

The particle size calculated from the decay rate of correlation curve with classical Einstein-Stokes equation \cite{Magatti2001} and the relative error for the four different particle diameters at four different scattering angles are between 1.2 and 5.7\% with an average deviation of 3\% as shown in Fig.6.

\section {Conclusions}

A key part of a photon correlator is a real-time correlation operation over a large span of lag time with ns sampling resolution. Among the most recent and best publications related to the FPGA correlator, we note that the multi-$\tau$ FPGA correlator proposed by Buchholz et.al. calculated the cross-correlation and autocorrelation simultaneously using a single avalanche diode array (SPAD) to detect photon pulses, although the minimum lag time starts only from 10 $\mu$s \cite{Buchholz2012}. The FPGA correlator proposed by Kalinin et al. was superior with a minimum lag time of 4 ns, but for input pulses sampled at 500 MHz, they used commercial photon counting cards \cite{Kalinin}.

 In this paper, we demonstrate an FPGA implementation of a digital correlator in which the lag time period reaches the range of 10 ns to 45 mins in real time. To our knowledge, this is the largest dynamic range ($10^{12}$) of an FPGA-based correlator implemented using the multi-$\tau$ technique. Consistently, we complement the correlator design with a fast and flexible counter for recording the detected photons in TCSPC mode at 800 MHz on the same FPGA device. With such a high speed, we are able to obtain the exact photon event arrival time. Then, we set the sampling time to 10 ns, in accordance with the requirements of the multi-$\tau$ technique, where it is greater than the input sample clock (1.25 ns) by eight times. Thus, we consider this correlator to be optimal in the sense of memory and computational accuracy. The overall design bitfiles and VHDL source code are available on \cite{DVN/TGEFXE_2019}.
 
 The proper functioning of the correlator is determined using aqueous suspensions of polystyrene spheres. Noise in the correlation function data is introduced through the detection of scattered light, and numerical methods are used to estimate the error by a calibration.
 Thus, polystyrene sphere diameters are determined with a deviation of 3\%, which can be considered an acceptable level of statistical error.

\section*{Acknowledgment}
This work was supported by the National Natural Science Foundation of China (Grant No. 61775065), and the Fundamental Reserach Funds for the Central Universities (HUST) under Grants 2016YXMS206.

\newpage
\begin{figure}[h]
\centerline{\includegraphics[width=115mm]{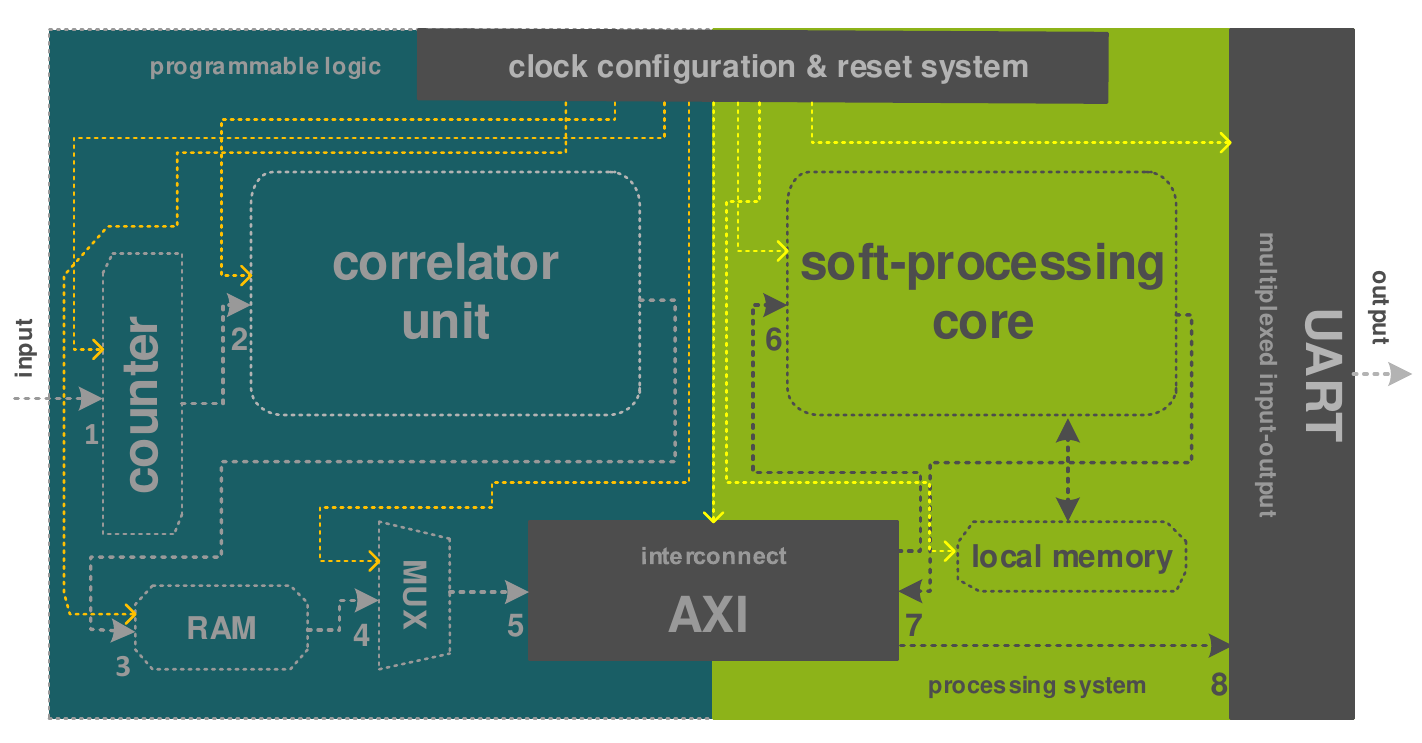}}
\caption{Schematic diagram of the correlation system based on the FPGA.}
\end{figure}

\begin{figure}[h]
\centerline{\includegraphics[width=140mm]{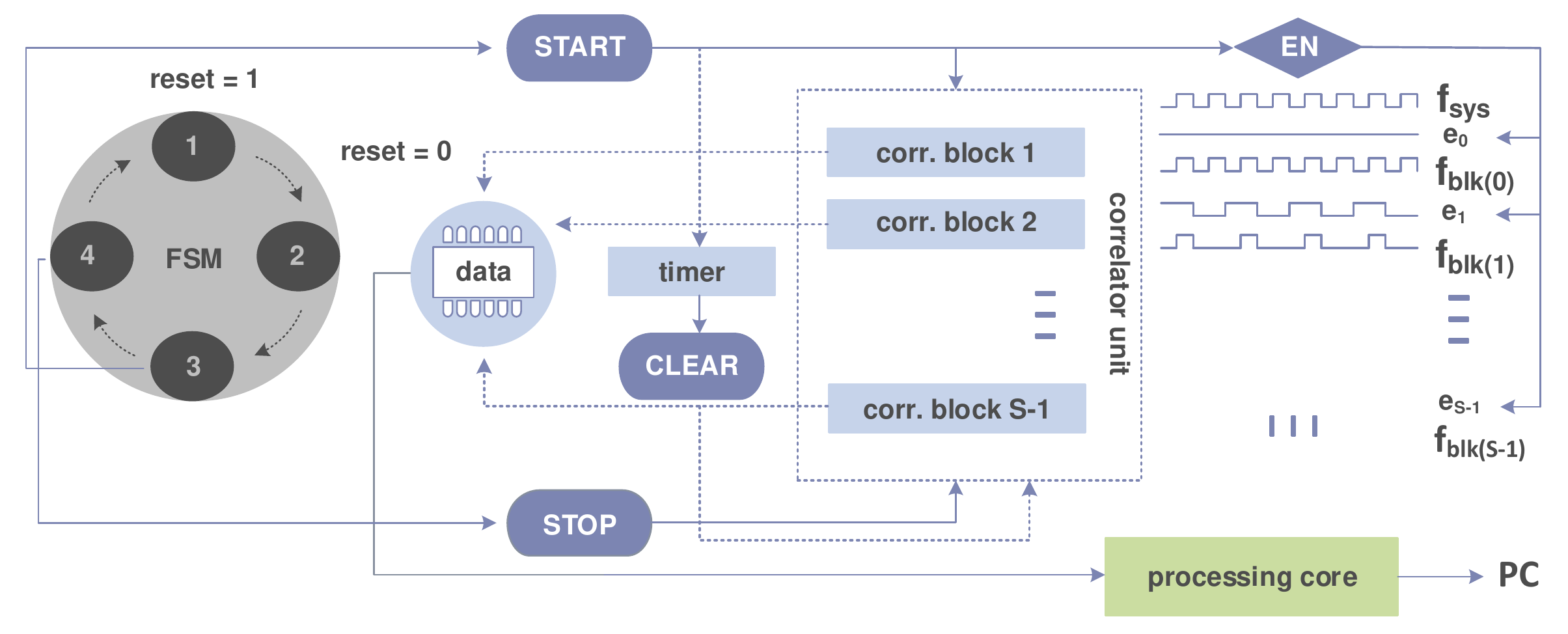}}
\caption{ Schematic diagram of the process control system.}
\end{figure}

\begin{figure}[h]
\centerline{\includegraphics[width=120mm]{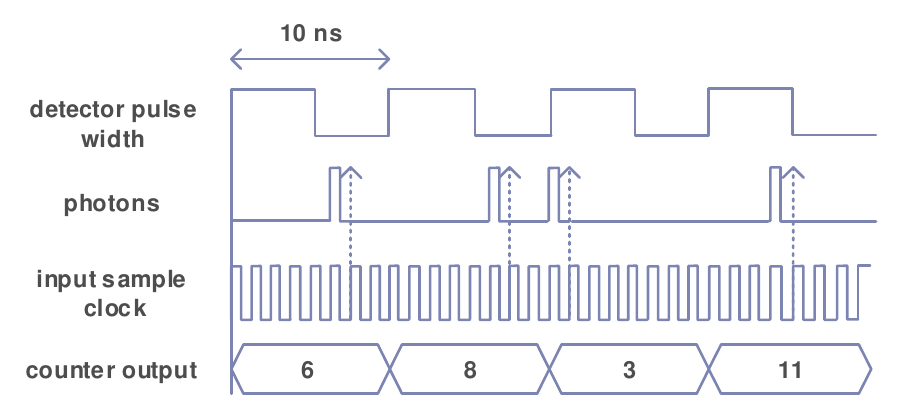}}
\caption{ Signal waveform demonstrating the working principle of the photon counter with the TCSPC technique.}
\end{figure}

\begin{figure}[h]
  \centering
     \subfigure a) {%
    \includegraphics[width=5 cm]{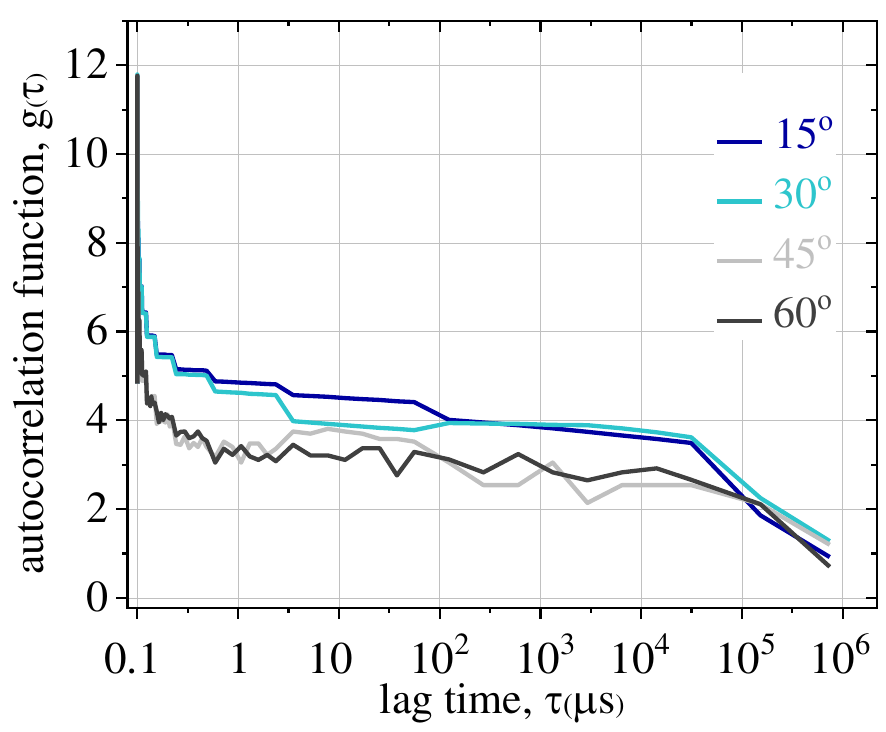}%
  }%  
    \subfigure b) {%
    \includegraphics[width=5.3 cm]{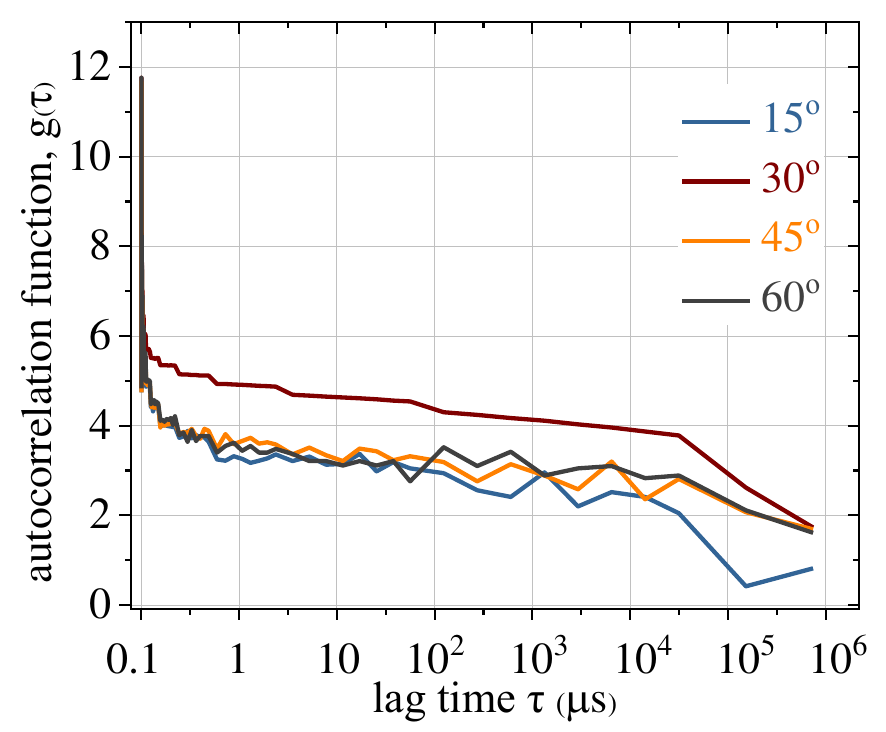}%
  }%  
\\
  \centering
     \subfigure c) {%
    \includegraphics[width=5.2 cm]{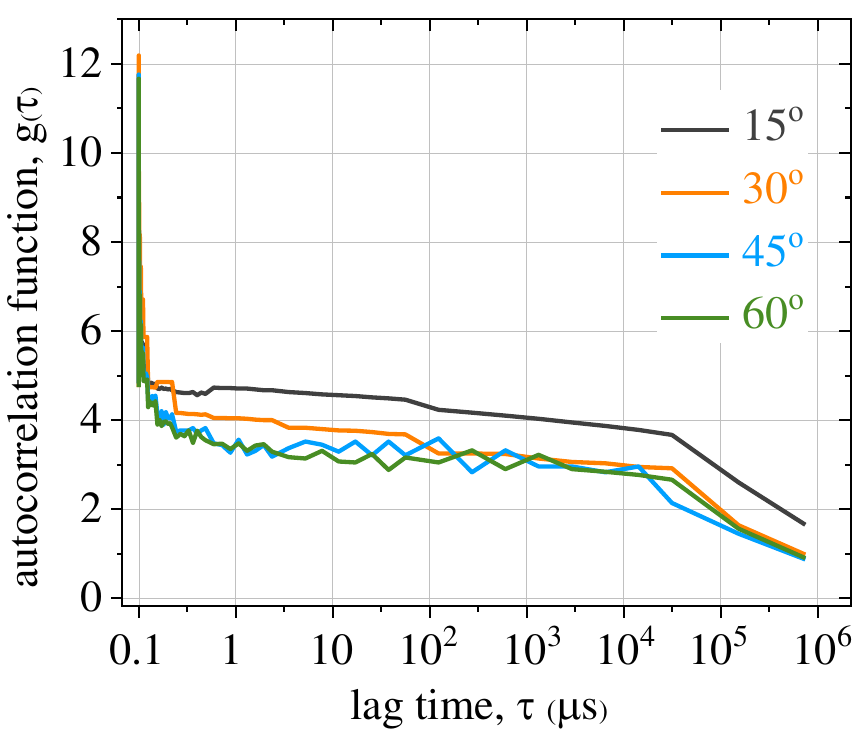}%
  }%  
    \subfigure d) {%
    \includegraphics[width=5.3 cm]{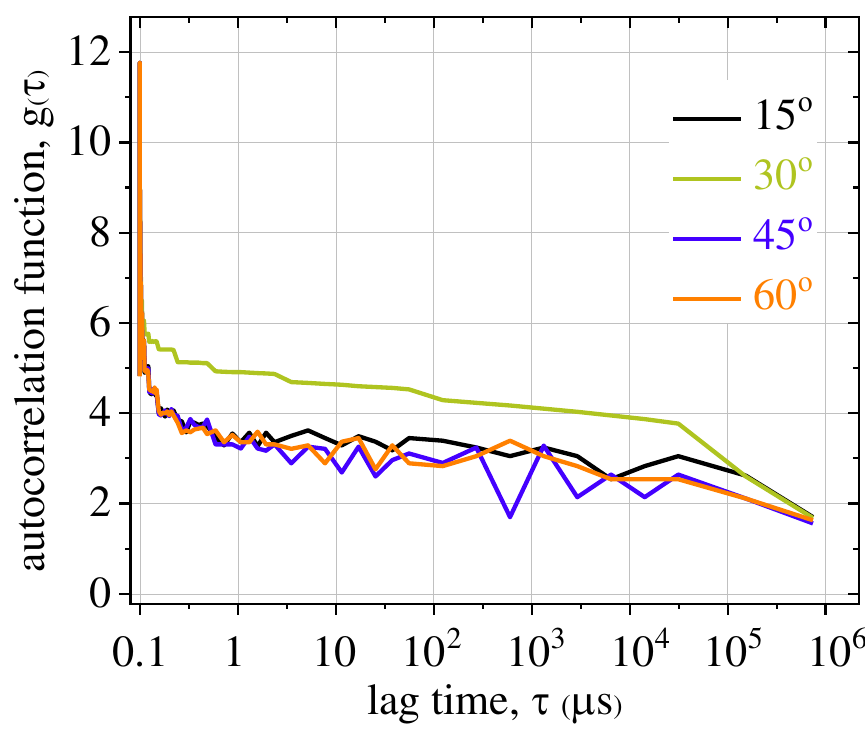}%
  }%  
  
  \caption{ Autocorrelation function of light scattering of particles with diameters of 530 nm (a), 805 nm (b), 240 nm (c), and 360 nm (d) at 4 scattering angles ($15^\circ$, $30^\circ$, $45^\circ$, and $60^\circ$)}
\end{figure}

\begin{figure}[h]
\centerline{\includegraphics[width=70mm]{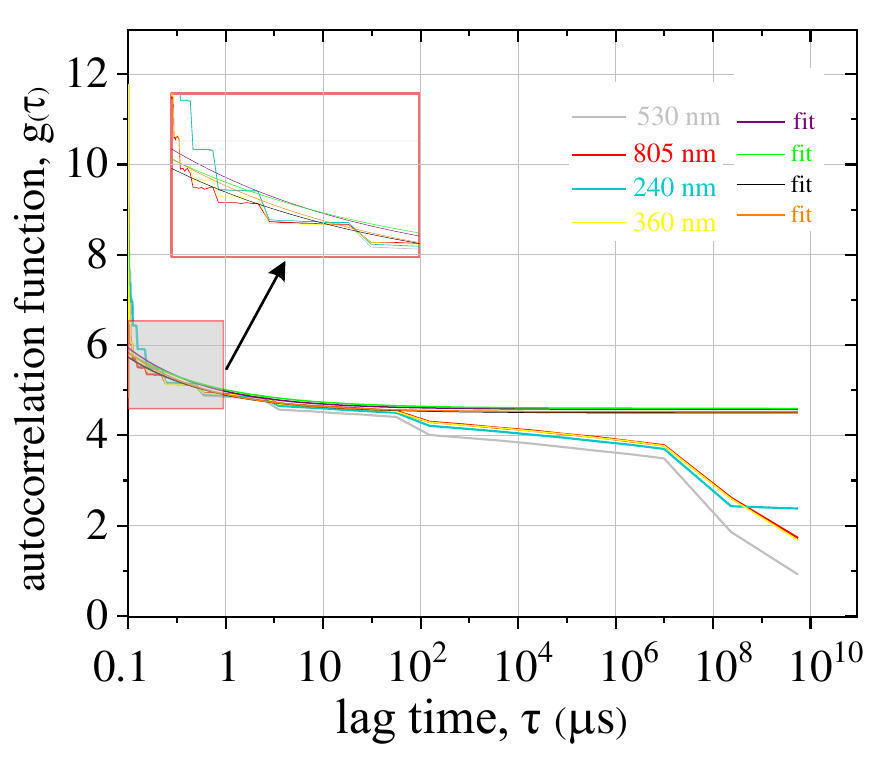}}
\caption{ The intensity correlation functions at a scattering angle of $30^\circ$ and their fits.}
\end{figure}

\begin{figure}[h]
\centerline{\includegraphics[width=75mm]{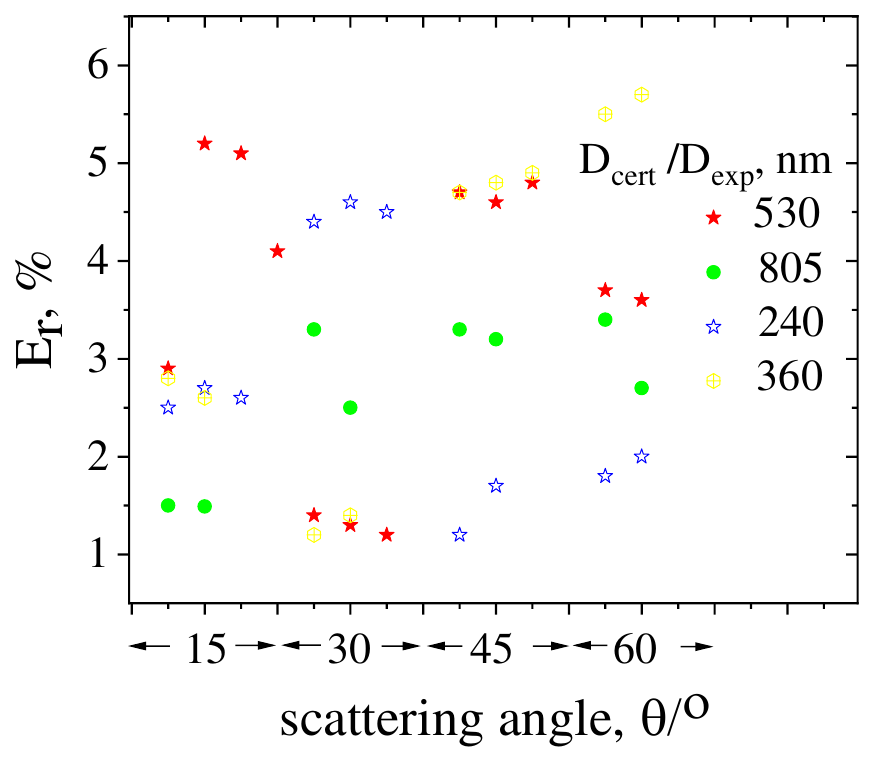}}
\caption{The ratio between the particle diameter calculated from the correlation curve \textit{$D_{exp}$} and the certified diameter \textit{$D_{cert}$} as a percentage of the relative error \textit{$E_{r}$} at different scattering angles $\theta$.}
\end{figure}


\section*{References}

\begin{thebibliography}{10}
\expandafter\ifx\csname url\endcsname\relax
  \def\url#1{\texttt{#1}}\fi
\expandafter\ifx\csname urlprefix\endcsname\relax\def\urlprefix{URL }\fi
\expandafter\ifx\csname href\endcsname\relax
  \def\href#1#2{#2} \def\path#1{#1}\fi

\bibitem{Engels99}
M.~Engels, B.~Hoppe, H.~Meuth, R.~Peters, A single chip 200 mhz digital
  correlation system for laser spectroscopy with 512 correlation channels,
  IEEE, Proceedings of the 1999 IEEE International Symposium on Circuits and
  Systems 5 (1999) 160--163.
\newblock \href {http://dx.doi.org/10.1109/ISCAS.1999.777535}
  {\path{doi:10.1109/ISCAS.1999.777535}}

\bibitem{Magatti2001}
D.~Magatti, F.~Ferri, Fast multi-tau real-time software correlator for dynamic
  light scattering, Appl. Opt 40 (2001) 4011--4021.
  \newblock \href {http://dx.doi.org/10.1109/ISCAS.1999.777535}
  {\path{doi:10.1364/ao.40.004011}}.

\bibitem{Magatti2003}
D.Magatti, F.~Ferri, 25 ns software correlator for photon and fluorescence
  correlation spectroscopy, Rev. Sci. Instrum. 74 (2003) 1135--1144.

\bibitem{Schaub2012}
E.~Schaub, F2cor: fast 2-stage correlation algorithm for fcs and dls, Opt.
  Expres. 20 (2012) 2184–2195.
\newblock \href {http://dx.doi.org/10.1364/OE.20.002184}
  {\path{doi:10.1364/OE.20.002184}}.

\bibitem{Jakob2007}
C.~Jakob, A.~T. Schwarzbacher, B.~Hoppe, R.~Peters, A fpga optimised digital
  real-time mutichannel correlator architecture, IEEE, 10th Euromicro
  Conference on Digital System Design Architectures, Methods and Tools 1 (2007)
  35--42.
\newblock \href {http://dx.doi.org/10.1109/dsd.2007.4341447}
  {\path{doi:10.1109/dsd.2007.4341447}}.

\bibitem{Buchholz2012}
J.Buchholz, J.W.Krieger, G.Mocsar, B.Kreith, E.Charbon, G.Vámosi,
  U.Kebschull, J.Langowski, Fpga implementation of 32x32 autocorrelator array
  for analysis of fast image series, Opt. Express 20 (2012) 17767--17782.
\newblock \href {http://dx.doi.org/https://doi.org/10.1364/OE.20.017767}
  {\path{doi:https://doi.org/10.1364/OE.20.017767}}.

\bibitem{Kalinin}
S.~Kalinin, R.~Kühnemuth, H.~Vardanyan, C.~A.~M. Seidel, Note: A 4 ns hardware
  photon correlator based on a general-purpose field programmable gate array
  development board implemented in a compact setup for fluorescence correlation
  spectroscopy, Rev. Sci. Instrum. (2012) 91--102.\href
  {http://dx.doi.org/10.1063/1.4753994} {\path{doi:10.1063/1.4753994}}.

\bibitem{Schatzel1985}
K.~Schatzel, New concepts in correlator design, Inst. Phys. Conf. Ser. 77
  (1985) 175--184.

\bibitem{Schatzel1988}
K.~Schatzel, M.~Drewel, S.~Stimac., Photon correlation measurements at large:
  lag times: Improving statistical accuracy, J. Mod. Opt. 35 (1988) 711--718.

\bibitem{Schatzel1990}
K.~Schatzel, Noise on photon correlation data: I. autocorrelation functions,
  Quantum Optics: Journal of the European Optical Society Part B 2 (1990)
  287--305.
\newblock \href {http://dx.doi.org/10.1088/0954-8998/2/4/002}
  {\path{doi:10.1088/0954-8998/2/4/002}}.

\bibitem{HUST}
L.~Li, L.~Yu, K.Yang, W.~Li, K.Li, M.Xia, Angular dependence of multiangle
  dynamic light scattering for particle size distribution inversion using a
  self-adapting regularization algorithm, Journal of Quantitative Spectroscopy
  and Radiative Transfer 209 (2018) 91--102.
\newblock \href {http://dx.doi.org/https://doi.org/10.1016/j.jqsrt.2018.01.022}
  {\path{doi:https://doi.org/10.1016/j.jqsrt.2018.01.022}}.

\bibitem{DVN/TGEFXE_2019}
A.~Islambek, \href{https://doi.org/10.7910/DVN/TGEFXE}{Fpga based digital
  correlator.}\href {http://dx.doi.org/10.7910/DVN/TGEFXE}
  {\path{doi:10.7910/DVN/TGEFXE}}.
\newline\urlprefix\url{https://doi.org/10.7910/DVN/TGEFXE}

\bibitem{Berne}
J.B. Berne, R. Pecora, Dynamic Light Scattering with Applications to Chemistry, Biology and Physics, Mineola, New York: 
  Courier Dover Publications (2000) 384.

\bibitem{Twomey}  
  S. Twomey, On the numerical solution of Fredholm integral equations of the first kind by the inversion of the linear system produced by quadrature, J. ACM 10 (1963) 97–-101.
 
 \bibitem{Buttgereit}  
  R. Buttgereit, T. Roths, J. Honerkamp, L. Aberle, Simultaneous regularization method for the determination of radius distributions from experimental multiangle correlation functions, Phys. Rev. E 64 (2001) 041404.
 
  \bibitem{Provencher}  
SW. Provencher, A constrained regularization method for inverting data represented by linear algebraic or integral equations, Comput. Phys. Commun. 27 (1982) 213-–27.

  \bibitem{ProvencherCON}  
SW. Provencher SW, CONTIN: a general purpose constrained regularization program for inverting noisy linear algebraic and integral equations, Comput. Phys. Commun. 27 (1982) 229-–42.

\bibitem{Clementi}
LA. Clementi, JR. Vega, LM. Gugliotta, HR. Orlande, A Bayesian inversion method for estimating the particle size distribution of latexes from multiangle dynamic light scattering measurements, Chemom. Intell. Lab. Syst. 107 (2011) 165-–73.

\bibitem{Gugliotta}
LM. Gugliotta, GS. Stegmayer, LA. Clementi, VDG. Gonzalez, RJ. Minari, JR. Leiza, JR. Vega, A neural network model for estimating the particle size distribution of dilute latex from multiangle dynamic light scattering measurements, Part. Part. Syst. Char. 26 (2009) 41-–52.

\end{thebibliography}
\end{document}